# SSL ENHANCEMENT

Fenil Kavathia
Student of Computer Engineering,
Nirma University.

A/29, Goyal park,PremchandNagar Road
Vastrapur,Ahmedabad-380015,India.
08bce028@nirmauni.ac.in

Ajay Modi
Student of Computer Engineering,
Nirma University.

4, Vivek Flat, Dashaporwad Soc. Paldi
Ahmedabad-380007, India.
mitajakom@gmail.com

## ABSTRACT
With the development of e-commerce, ssl protocol is more and more widely applied to various network services. It is one of key technologies to keep user's data in secure transmission via internet.

This document majorly focuses on sslstrip which generates the most recent attack in the secure network connections. It strips out all the secure connections to unsecure plain connection.

In this article we depict this attack and to nullify it, we have proposed a technique cum practical solution to strengthen data security by developing mozilla-firefox add-on and servlet code which will strengthen our defense against the https hijacking attacks.

## General Terms
This paper depicts a SSL breach and then provides a solution to nullify it. So this paper comes under Network security field.

## SSL PROTOCOL

### Motivation
With the development of e-commerce, ssl protocol is more and more widely applied to various network services. Ssl protocol can provide end to end authentication, message encryption, message integrity and other security mechanisms to protect the security of the communications. However, in practical applications, ssl is not flawless; there may be possibility of man-in-the-middle attack. Many researches focus on https hijacking and its defense. In 2003, peter burkholder analyzed ssl handshake defect and verify the possibility of attack [1] to ssl. In 2009, Michael Howard in his article described conducting ssl attacks by webmitm [2]. On international security conference in 2009, moxie declared that in practical applications, the redirection from http to https connections would be security risk [3]. Reference [4] [5] [6] describe the process of https attacks and its protection. This paper analyzes the where exactly client fails to have an end to end secure connection and verifies the hijacking attack to https. After analyzing the sslstrip attack, we have proposed prevention schemes to strengthen network security.

### Structure Of SSL
The ssl protocol includes two sub-protocols: the ssl record protocol and the ssl handshake protocol. The ssl record protocol defines the format used to transmit data. The ssl handshake protocol involves using the ssl record protocol to exchange a series of messages between an ssl-enabled server and an ssl-enabled client when they first establish an ssl connection. This exchange of messages is designed to facilitate the following actions:

- Authenticate the server to the client.
- Allow the client and server to select the cryptographic algorithms, or ciphers, that they both support.
- Optionally authenticate the client to the server.
- Use public-key encryption techniques to generate shared secrets.
- Establish an encrypted ssl connection.

Here, we give the illustration of a full hand-shake protocol.

### *The ssl handshake*
The ssl protocol uses a combination of public-key and symmetric key encryption. Symmetric key encryption is much faster than public-key encryption, but public-key encryption provides better authentication techniques. An ssl session always begins with an exchange of messages called the *ssl handshake*. The handshake allows the server to authenticate itself to the client using public-key techniques, and then allows the client and the server to cooperate in the creation of symmetric keys used for rapid encryption, decryption, and tamper detection during the session that follows. Optionally, the handshake also allows the client to authenticate itself to the server.

The exact programmatic details of the messages exchanged during the ssl handshake are beyond the scope of this document. Reference [8] describes the process of full handshake of a protocol.

## Where Exactly Client Fails:

### Https connection initiates by http:
According to analysis of user's habits and the practical applications of HTTPS, HTTPS request will be initiated by the following two ways:

### Users' habits
When user accesses HTTPS sites using web browser, it usually type directly the URL without https head in the address bar, such as: www.xxx.com. If there is no protocol head in URL, browser will use the HTTP protocol to connect the site. When client initiates HTTP connections, but the server is the HTTPS site, it will return messages of HTTP redirection. The content in this packet contains the actual HTTPS address, such as https://www.xxx.com. The client receives this packet, and browser will be re-launched to initiates HTTPS connection. Compared with directly typing https://www.xxx.com in client browser, this redirection will be no difference.

### Application in practice
With HTTP connection, there is some button on the page to initiate HTTPS connections. For example, when you want log in personal account of E-mail, you will click on the submit button to transmit your ID and password. When you click on the button, the client initiates HTTPS connections to protect confidentiality of personal information.

**Case a) (1.3.1.1)** is because of the habits of users, they don't pay attention to the difference between http and https in the URL.

**Case b) (1.4.1.2)** is because of consideration of the overhead of SSL handshake, only important information has encrypted, not the whole data in the connection. In general, websites don't use HTTPS connection in the whole process, because HTTPS connection is usually 2 to 100 times slower than HTTP connection s. Therefore, submission of confidential information (such as ID, password) is by HTTPS connection, and other services are still using HTTP connection. In this way, the delivery of HTTPS URL is in the HTTP message content, while the inherent insecurity of HTTP protocol, so it results in security vulnerabilities.

## Who Says Https Is Secure?
In this section we are particularly focuses on functioning of SSLSTRIP tool which strips out secure connection and force a client to initiate an unsecure connection.

### SSL Strip-A disastrous Attack on https
content, and sends to the client. At the same time, attacker initiates HTTPS handshake with the server. If attacker receives the HTTP request from client which receives the tampered content, it decrypts the data in cipher text receiving from server and sends it to the client. Because of HTTP connection between client and attacker, there is no alert dialog in the browser of client. So this method is better deceptive**.** Attack has following steps**:**

Figure 1 explains how attacker works as an intruder using sslstrip tool and how he/she maintains two different connections.

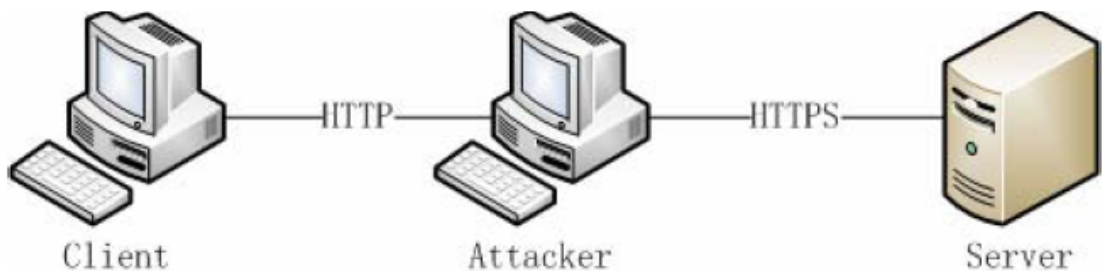

**Fig. 1 Sslstrip attacks on secure connection**

*a)* Conducting ARP Spoofing, it can make attacker in the middle of server and client. Then attacker can forward packets between server and client.

*b)* Attacker monitors the HTTP data between server and client.

*c)* When attacker receives the data content from server has <a href="https://...">, replacing it with <a href="http://...">. If the head of HTTP packet has "location: https://…", replacing it with "location: http://…", making a record about the addresses which have tampered before.

Figure 2. explain functioning of the sslstrip attack and how it strip secure connection of a victim.

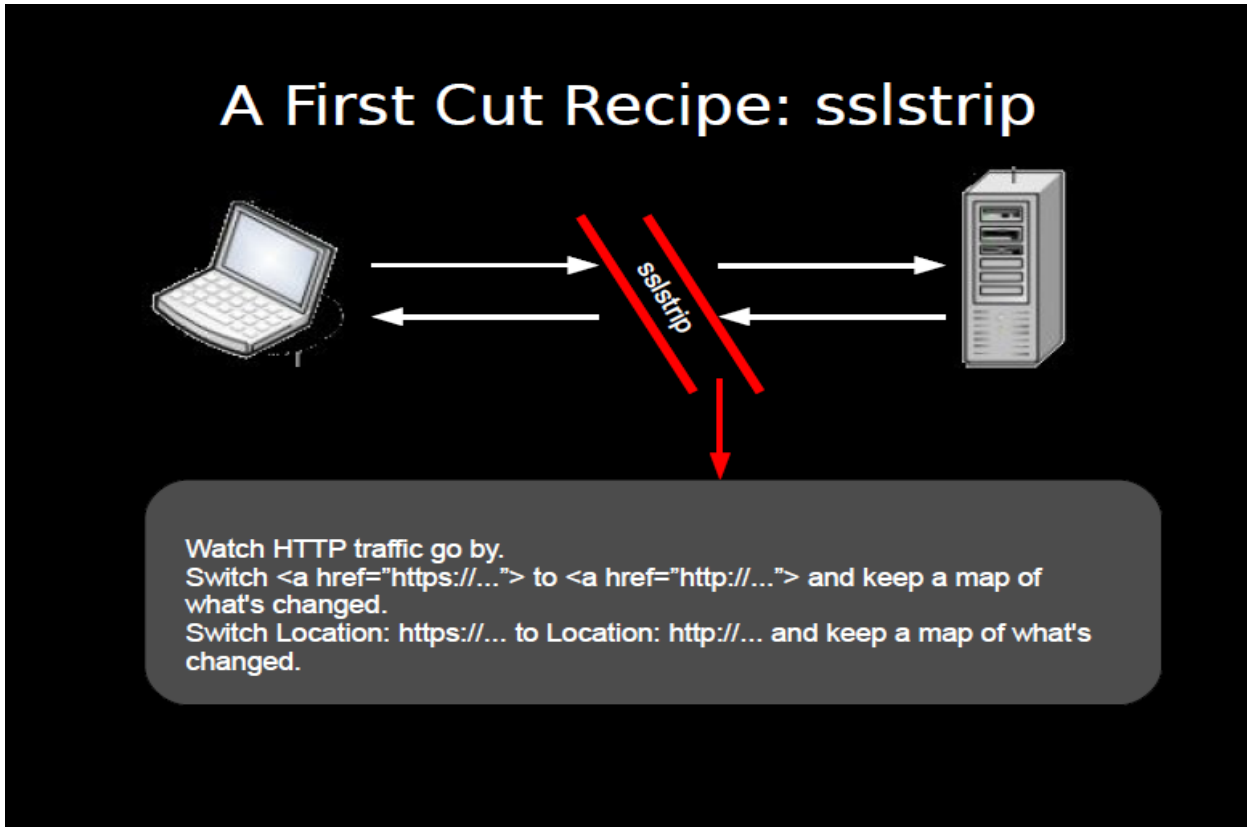

**Fig. 2 Sslstrip Functioning.**

*d)* If attacker receives the HTTP request to the address which is in the record, it connects with the server by HTTPS.

*e)* After establishing HTTP connection with client, attacker decrypts the cipher text of HTTP from the server, and sends the plaintext of HTTP to the client. In the browser of client, there is no difference with the normal HTTPS connection, except the URL. In normal way, the URL begins with https. But after attacking, it begins with http[7].

## 1.6 Experimental Result

Experiments are conducted in 100M Ethernet, and the browser of client is Firefox 3.0. Client machine runs on windows xp, which IP address is 192.168.1.2. Attacking machine runs on Linux whose kernel is 2.6.11, and IP address is 192.168.1.3.

### 1.6.1 The result of SSL attack using sslstrip

The same program we use above to carry out ARP Spoofing and DNS Spoofing. With OpenSSL library and sslstrip of Moixe , we carry out attack. The result is shown below:

There is a significant difference in this picture. The norm is that the most websites begin with https, but after exploiting vulnerability it is http; the other one is that normal connection will have a lock icon under the browser, but after attacking there is not[3]. Figure 3 will help you to understand it clearly.

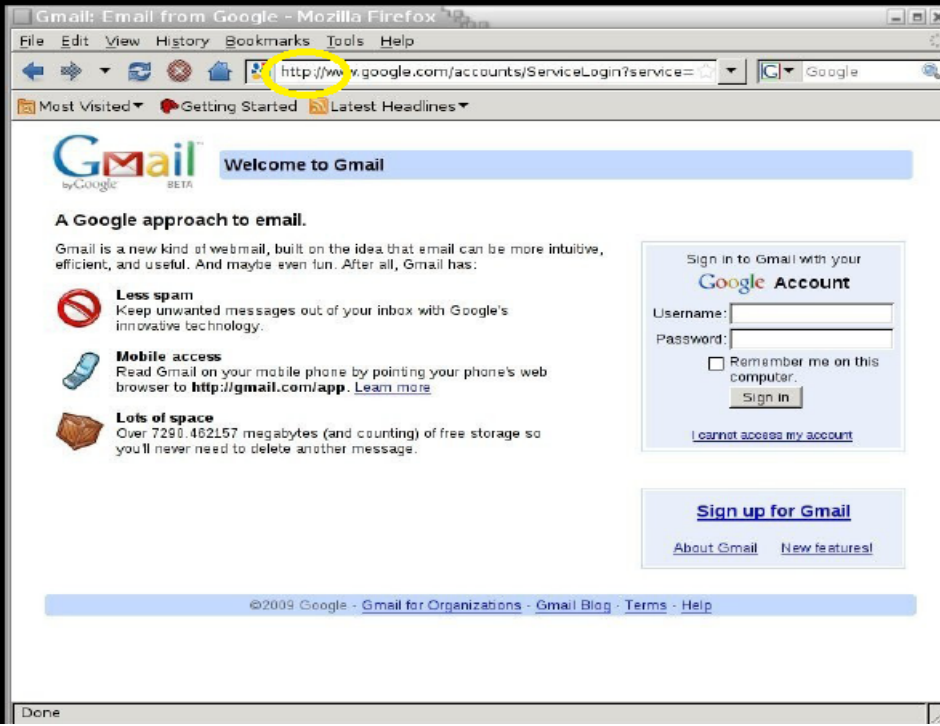

**Fig 3 Result of the sslstrip attack.**

## PREVENTION

We have seen there's no negative alert feedback approach from the browser as attacker is maintaining http connection between client and attacker and usual authentic connection between attacker and server that is usually in https. So server wouldn't be able to find out whether anything fraudulent activity working in between or not. So what we can do now?

What if a client strictly tells that he/she want secure connection for a particular site if available. If a client will aware about initiating a secure connection from his machine then intermediate attacker will surely going to fail. So we have thought that we will make a firefox add-on particular for one browser Mozilla firefox which will dynamically send URL request or rather say server host/ip request to a server.

## STEP BY STEP HOW ADD-ON EXACTLY WORKS?

Figure 4 shows the graphical view of our add-on and its functioning.

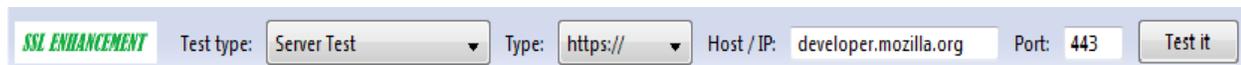

**Fig. 4 Suggested Add-on graphical view.**

Firstly user enters a URL in the URL bar of a browser on which he/she redirects to the link. But if attacker works as an intruder in between, then rather than redirecting to https connection user may end up at http connection which is a non-secure version. The main thing here to note is that even if application provider and client both willing to establish a connection in a secure mode, intermediate attacker wouldn't allow them to do so.

Our add-on provides solution of it. It dynamically fetches the URL from the URL bar to which user is redirected. Hence we are able to detect many things like Who is the host/IP ,what are the ports on which client and server communicates and several other information regarding the same.

Now on getting the information about what user actually got, from what he has requested, we will send the fetched URL from the victim's URL bar and strip this URL and make it convenient to a server request. We are fetching URL and determine whether connection is http/https/ftp according to that we will set port no 80,443 or 21.

 after getting a convenient URL we passed it to servlet. A servlet which will be running on the machine of a victim or on some domain which support java servlet and jsp programs. Now servlet on getting URL will first convert host/ip to Inetaddress and then will perform port scanning. Generally http request came on port 80 and https request came on 443 port. So to determine whether victim's request is actually stripped by sslstrip or not we are sending the server request on both the port as on 80 and on 443 and each time while getting the response on the same port for the same site we are incrementing the counter for the same.

So if we are able to receive a response on 80 port hence we are able to get response for http. Hence increment a counter by 1. If a particular site also responded on port 443 then we can say that it is having http and https both the connections available. Hence counter becomes 2 in that case. If a counter remains 0 simply means you didn't have a valid URL or maybe you are simply not connected to server.

 The main thing is to be noted here is that we are performing it in a dynamic way hence if a site is not available in https at any cost than at runtime we are not able to get any response at port 443. Hence we can say our site is actually not available in https and just available in http.
What actually sslstrip does **is stripping all urls , all links and all cookies which are available in https and send a victim http links**. So when we are forcing our links to be remain in the https by using port scanning we help a victim to send encrypted data over the channel and not plain text which anyone can easily get. But cipher text can't be easily decrypted as our browser is using rc5, md5 and sha1 such algorithms which provide high level encryption which can't be easily decrypted. Following the same way we are providing security to our client.

So when we are able to scan a site which is available on port 443 Hence servlet will send a new request to browser that will help to switch over to https for the same site into the new tab. Hence whenever a site is available in https and we are redirected to http we strictly switch over to https which prevent the attack of sslstrip.
In the given below figure 5 you can clearly distinguish that how our add-on help a client to maintain end-to-end secure connection.
Also figure 6 shows how our solution save client to reveal his password to the attacker and secure safety of his account.

We are just showing one application of our solution. It can use for multiple purpose, for ensuring & enhancing network security.

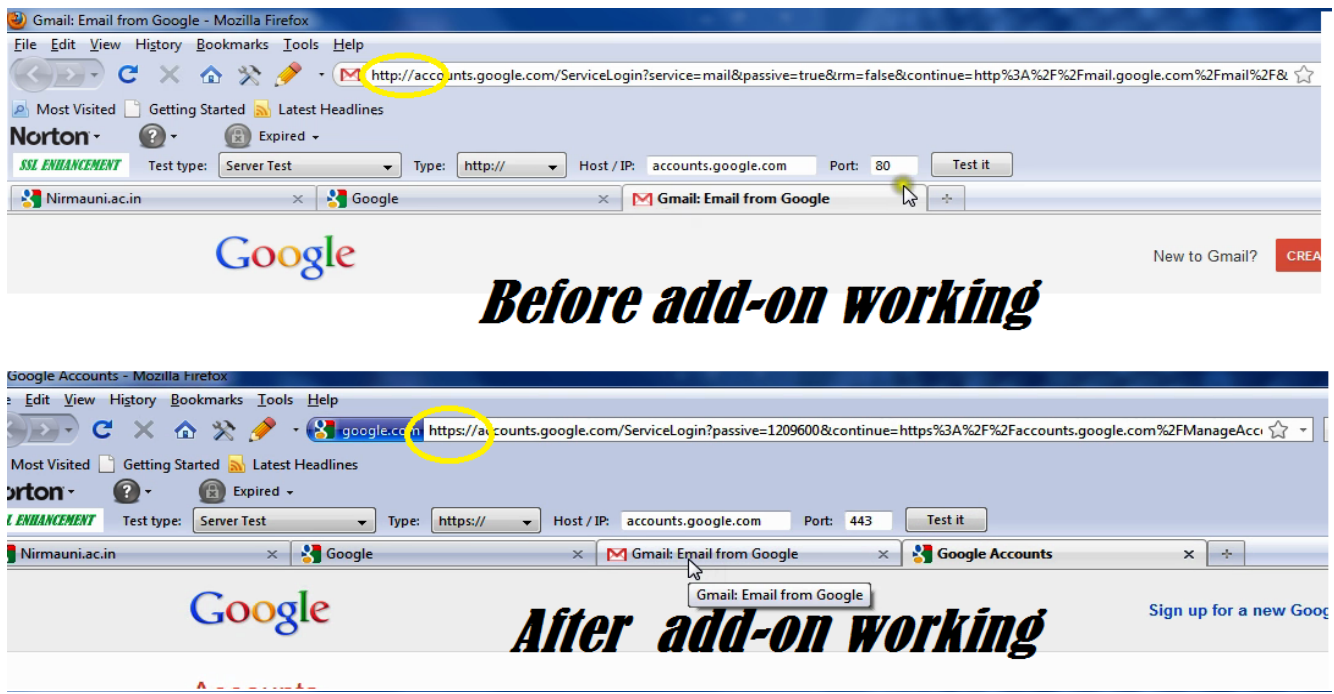

**Fig. 5 Before and After add-on working.**

```
2011-10-20 10:39:22,142 SECURE POST Data (accounts.google.com):
continue=http%3A%2F%2Fmail.google.com%2Fmail%2F&service=mail&rm=false
&dsh=6250000089884133564<mpl=default<mpl=default&scc=1&GALX=JmZl2Lhz4UE&pstMsg=1&dnConn=&timeStmp=&secTok=
&Email=08bcexxx&Passwd=08BCEXXX&signIn=Sign+in&PersistentCookie=yes&rmShown=1
```

**Before Add-on working password stripped by attacker**

```
2011-10-20 10:43:09,933 POST Data (safebrowsing.clients.google.com):
goog-malware-shavar;a:38160-53307:s:50810-63942:mac
goog-phish-shavar;a:159187-170524:s:78575-82826:mac

2011-10-20 10:45:29,558 POST Data (ocsp.thawte.com):
0q0o0M0K0I0 ACK ENQ + SO ETX STX SUB ENQ  EOT DC4 RS '  ªq<yKÊ RS " SUB
a-?Ð°`ƒ EOT DC4 ;4šp's² Š ESC FF óé7Í³p2ž CAN T STX DLE #…d)!"€ RS a‰ÄQ¢tû÷¢ RS 0 FS 0 SUB ACK   + ACK SOH ENQ ENQ BEL 0 SOH EOT EOT
0 VT ACK   + ACK SOH ENQ ENQ BEL 0 SOH SOH
2011-10-20 10:45:31,491 POST Data (ocsp.thawte.com):
0q0o0M0K0I0 ACK ENQ + SO ETX STX SUB ENQ  EOT DC4 RS '  ªq<yKÊ RS " SUB
a-?Ð°`ƒ EOT DC4 ;4šp's² Š ESC FF óé7Í³p2ž CAN T STX DLE /ß¼ö® 'Rm SI š£ß@4>š¢ RS 0 FS 0 SUB ACK   + ACK SOH ENQ ENQ BEL 0 SOH EOT EOT
0 VT ACK   + ACK SOH ENQ ENQ BEL 0 SOH SOH
```

**After Add-on working password encrypted and can't stripped**

Fig 6. Our solution fails sslstrip attack.

**************************************

## ACKNOWLEDGMENTS

Our sincere thanks to the Prof. Anitha Ashokan,( Nirma Institute of Technology, India) who have guided throughout our work.

## REFERENCES

[1] Peter Burkholder, "SSL Man-in-the-Middle Attacks", SANS Institute InfoSec Reading, 2003.

[2] Michael Howard, "Man-in-the-Middle Attack to the HTTPS Protocol", IEEE computer society, 2009, pp.78-81.

[3] Marlingspike Moixe, "New Tricks For Defeating SSL in Practice", BlackHat Conference, USA(2009).

[4] Haidong Xia and Jose Carlos Brustolonl, "Hardening Web Browsers Against Man-in-the-Middle and Eavesdropping Attacks". Proc. 14th Int'l Conf. World Wide Web(IW3C2), ACM Press, 2005, pp. 489-498.

[5] Thawatchai Chomsiri, "HTTPS Hacking Protection", 21st International Conference on Adcanced Information Networking and Applications Workshops(AINAW), 2007.

[6] Andre Adelsbach and ebatian Gajek, "Visual Spoofinging of SSL Protected Web Sites and Effective Countermeasures", Proceedings of the 1st Information Security Practice and Experience Conference, Singapore, 11-14 April, 2005.

[7] Moxie,Ssltrip. http://www.thoughtcrime.org/software/sslstrip/.

[8]http://www.mozilla.org/projects/security/pki/nss/ssl/traces/trc-clnt-ex.html   A  SSL research    group  owned by mozila.